# Can Google Scholar and Mendeley help to assess the scholarly impacts of dissertations?[1]

Kayvan Kousha and Mike Thelwall

Statistical Cybermetrics Research Group, School of Mathematics and Computer Science, University of Wolverhampton, Wulfruna Street, Wolverhampton WV1 1LY, UK.

E-mail: {k.kousha, m.thelwall}@wlv.ac.uk

Dissertations can be the single most important scholarly outputs of junior researchers. Whilst sets of journal articles are often evaluated with the help of citation counts from the Web of Science or Scopus, these do not index dissertations and so their impact is hard to assess. In response, this article introduces a new multistage method to extract Google Scholar citation counts for large collections of dissertations from repositories indexed by Google. The method was used to extract Google Scholar citation counts for 77,884 American doctoral dissertations from 2013-2017 via ProQuest, with a precision of over 95%. Some ProQuest dissertations that were dual indexed with other repositories could not be retrieved with ProQuest-specific searches but could be found with Google Scholar searches of the other repositories. The Google Scholar citation counts were then compared with Mendeley reader counts, a known source of scholarly-like impact data. A fifth of the dissertations had at least one citation recorded in Google Scholar and slightly fewer had at least one Mendeley reader. Based on numerical comparisons, the Mendeley reader counts seem to be more useful for impact assessment purposes for dissertations that are less than two years old, whilst Google Scholar citations are more useful for older dissertations, especially in social sciences, arts and humanities. Google Scholar citation counts may reflect a more scholarly type of impact than that of Mendeley reader counts because dissertations attract a substantial minority of their citations from other dissertations. In summary, the new method now makes it possible for research funders, institutions and others to systematically evaluate the impact of dissertations, although additional Google Scholar queries for other online repositories are needed to ensure comprehensive coverage.

## 1 Introduction

Doctoral dissertations are important single-authored scholarly works written by early career researchers and form a significant minority of the scientific output of universities. For instance, according to the *British Library EThOS service* (https://ethos.bl.uk/), 1,114 doctoral theses were awarded by the University of Cambridge in 2015, in comparison to 8,174 Scopus-indexed journal articles. Doctoral theses usually include a comprehensive literature review, detailed original findings, a discussion, or another significant contribution to scholarship based upon three or more years of full-time equivalent research. They are assessed by independent examiners before being published, and so could be described as peer-reviewed. Dissertations may make substantial contributions to scholarship or professional practice in some fields, such as for supporting clinical practice (McLeod & Weisz, 2004). Citations to dissertations may also be important impact evidence for early career researchers since they may have had too little time to have journal articles published (ACUMEN Consortium, 2014). This would be particularly relevant for job applications but may also impact early promotion decisions or grant applications. Moreover, many

---

universities, departments, research committees or funders of doctoral research may wish to monitor the success of their doctoral programs, including their scientific, social, economic, clinical or cultural benefits, requiring alternative metrics for the impact assessment of non-standard academic outputs (Kousha & Thelwall, 2015; Schöpfel & Prost, 2016).

Although peer review is the best method to assess the quality of theses within a doctoral programme, they may be too large to be read by recruiters, promotion committees or funding agencies. Hence, many researchers have examined the publication rates generated from doctoral dissertations as an indication of doctoral program success or productivity (e.g., Thomas & Reeve, 2006; Lee, 2000; Stewart, Roberts & Roy, 2007; Echeverria, Stuart, & Blanke, 2015). The articles published from a dissertation don't necessarily reflect all its impact, however (Morse, 2005). Citation analysis could be useful for dissertations in theory but they are absent from the traditional citation indexes, such as the Web of Science and Scopus. Alternatively, counting citations to publications resulting from dissertations may be used to estimate their scientific impact (Larivière, 2012). Nevertheless, many doctoral students do not produce publications from their dissertations even several years after graduation (e.g., Anwar, 2004; Caan & Cole, 2012; Evans et al., 2018). For instance, there is large-scale evidence that most doctoral students in the arts and humanities (96%) and in social sciences (90%) had no Web of Science publications during 2000-2007 (Larivière, 2012). Moreover, it might be difficult to correctly track publications from dissertations several years after graduation and to assign authorship credit for multi-authored publications resulting from dissertations (Hagen, 2010). Given the lack of a comprehensive citation index, cited references searches in conventional citation indexes have also been used to assess citations to theses or dissertations by searching for related terms (e.g., *thesis* or *dissertation*) in the reference sections of other scholarly publications (Larivière, Zuccala, & Archambault, 2008; Rasuli, Schöpfel, & Prost, 2018). This method is useful to estimate the total number of citations to all dissertations rather than to individual doctoral dissertations across different fields, institutions or years. Moreover, using conventional citation databases for this purpose could be problematic in the social sciences and arts and humanities, where poorly-indexed books and publications in languages other than English could be important (Archambault et al., 2006; Huang & Chang, 2008).

Google Scholar is not primarily a citation index for dissertations, but automatically indexes dissertations from many institutional repositories, databases and commercial publisher websites and reports counts of citations to them based on its indexed publications. By November 2018, Google Scholar had indexed the metadata or text of 264,000[2] UK doctoral dissertations from *The British Library EThOS* service, which is the UK's national database of doctoral theses (https://ethos.bl.uk/) and 323,000[3] French Ph.D. theses from *theses.fr*. Google Scholar has also indexed many dissertations from university or institutional repositories, making it possible to assess the citation impact of dissertations at institutional level. For example, *the University of Glasgow* (http://theses.gla.ac.uk/) and *London School of Economics and Political Science* (http://etheses.lse.ac.uk/) have repositories for postgraduate theses which have been mostly indexed by Google Scholar (est. 6,650 of 7,372[4] and 3,560 of 3,691[5] respectively). Moreover, 56% (1,991 of 3,520) of repositories indexed by the *Directory of Open Access Repositories* (OpenDOAR, www.opendoar.org/) had theses or dissertations subsets in different countries. Nevertheless, many dissertations might not be searchable by commercial search engines or digital libraries due to copyright, publication embargos, submission of dissertations in print format, restricted access, or non-searchable websites (Kettler, 2016). For instance, although *DART-Europe E-theses Portal* (http://www.dart-

---

2 This is an estimation based on searching the *site:ethos.bl.uk* command in Google Scholar query
3  This is an estimation based on searching the *site:theses.fr* command in Google Scholar query
4 This is an estimation based on searching the *site:theses.gla.ac.uk* command in Google Scholar query
5 *This is an estimation based on searching the site:etheses.lse.ac.uk command in Google Scholar query*

europe.eu/) claims to include "805,570 open access research theses from 617 Universities in 28 European countries", Google Scholar has not directly indexed any from this portal.

*ProQuest Dissertations & Theses* is a digital library that indexes and provides full-text access to dissertations and theses. It claims to include "2 million full text dissertations" from more than "3,000 schools", mostly North American universities[6]. In October 2017, ProQuest announced that the contents of "half a million dissertations" would be indexed by Google Scholar, linking users via their library ProQuest subscription. Unsubscribed users can also usually "access the first 24 pages at no charge" (Arbor, 2017). In November 2018, Google Scholar had indexed about 250,000 records from the *ProQuest Dissertation & Theses*[7].

Although a few studies have manually analysed Google Scholar citations to small numbers of dissertations (see below), there have been no large-scale multidisciplinary citation assessments of doctoral dissertations. This is a regrettable omission, given the potential value of dissertations for early career researchers. The current study fills this gap by testing a method to systematically extract Google Scholar citations to 77,884 ProQuest-indexed doctoral dissertations from 2013-2017 across 18 fields. Mendeley reader counts for all doctoral dissertations were also collected for comparison with citation counts.

## 2  Background

### 2.1  References in dissertations

Previous studies have investigated references in dissertations mainly to identity the cited publication types, the core serials cited or the citation practices of students in different subject areas (e.g., Haycock, 2013; Barnett-Ellis & Tang, 2016; Yeap & Kiran, 2017). These have all been small-scale studies, but suggest that few of the references in a dissertation are other dissertations and that there are disciplinary differences in this proportion.

An analysis of references from 49 management doctoral dissertations (2004 to 2009) found that most cited references were journals (65%) and books (19%), with less than 1% being other theses (Kumar & Dora, 2011). A study of 124 Ph.D. dissertations found some disciplinary differences in types of cited sources. For instance, in most science fields (e.g., Botany and Microbiology, Chemistry, Mathematics, Geology) journal articles were cited most frequently (56%-73%), whereas in Archaeology and Anthropology books were more important (40%) followed by journal articles (35%) and dissertations (7%). In Computer Science, journal articles (31%) and conference papers (18%) were most often cited in doctoral research (Salami & Olatokun, 2018). A study of 30 chemical sciences Ph.D. theses also found that journals (79%) were dominant, followed by books (16%) and only 0.2% of the citations were to other dissertations (Gohain & Saikia, 2014). In contrast, a study of 101 master's theses and doctoral dissertations (2005-2014) in engineering fields found that papers (27%), dissertations (20%) and reports (17%) were cited most often (Becker & Chiware, 2015) and another study of 70 postgraduate dissertations in education (1992-2002) showed that books and monographs were most commonly cited (60%) followed by journal articles (25%) and theses (5%) (Okiy, 2003). Thus, it is clear that the types of cited references in theses and dissertations vary between subject areas.

### 2.2  Citations to dissertations

In the absence of a citation index for dissertations, the cited references search facilities in traditional citation indexes have been used to estimate the citations to theses or dissertations. Thomson Scientific's citation databases (now Clarivate Analytics) has been used to identify citations from articles, research notes and review articles to theses. These citations were identified by first extracting references matching the query thesis* and then using a set of rules to filter out non-thesis references (Larivière, Zuccala, & Archambault, 2008). This study found that the share of references to dissertations from articles, research notes and review articles declined the last century (1900-2004) including in the most recent period and in all broad fields examined. The decline might be due to scholars preferring to cite journal articles, conference papers or books generated from theses, rather than the original dissertations. The introduction of electronic theses has not reversed this decline (Larivière, Zuccala, & Archambault, 2008). This study did not examine the extent of citations to theses from non-journal documents and has an unknown coverage (recall) of theses.

Using a long manually-curated Scopus cited reference search in July 2018 to find references from Scopus-indexed documents to theses (REFSRCTITLE("(Dissertation)" OR "[dissertation]" … OR "M.S. thesis" …), another study found that about 1.5% of Scopus publications had at least one citation to a dissertation (Ph.D. dissertations: 69%; master's: 21%; other theses: 10%). In partial contrast to the above study, this found that the percentage of documents with at least one citation to a dissertation had increased 1996-2018 in each of 4 broad areas (Medical Sciences, Science & Technology, Social Sciences, Arts & Humanities) (Rasuli, Schöpfel, & Prost, 2018).

Combining the results of the above two studies, it is possible that a greater proportion of scholarly outputs cite theses but these citations form a smaller fraction of all citations, with citation lists expanding over the same period. These studies also show that it is possible to extract information about citations to theses from the Web of Science and Scopus but this information does not include citations from other dissertations and most books and seems to be more useful to assess overall citation pattern to all dissertations rather than to individual theses. Moreover, neither method could guarantee comprehensive coverage of any collection of dissertations. Google Scholar has the potential to address at least the former issue since it indexes dissertations, books and preprints.

Previous Google Scholar studies have shown that it can locate citations to non-standard publications, such as gray literature (Orduna-Malea, Martín-Martín, & López-Cózar, 2017) and books (Kousha, Thelwall, & Rezaie, 2011). It is helpful for research evaluation of humanities and social sciences programs (Prins, Costas, van Leeuwen, & Wouters, 2016; Halevi, Moed, & Bar-Ilan, 2017) by locating many citations from non-journal sources (48%-65%), such as books, conference papers, theses, and unpublished materials (Martín-Martín, Orduna-Malea, Thelwall, & López-Cózar, 2018).

A few studies have assessed the citation impact of small sets of individual dissertations based on Google Scholar citations. One study found a low correlation between downloads and citations to 16 digitized London School of Economics theses with at least ten citations, although the sample size was too small for meaningful results. The study also showed that the ten most downloaded dissertations from the ProQuest digitization project did not necessarily have the most Google Scholar citations (Bennett & Flanagan, 2016). Another study used Google Scholar citations to assess scientific impact of 97 South African doctoral dissertations in educational sciences from 2008, finding that 76% had no citations (including self-citations) from other publications indexed by Google Scholar (Wolhuter, 2015). Google Scholar citations to 125 Ph.D. theses in five broad subjects (25 theses per field) from the University of Salamanca were more common in Experimental Sciences (32%), Social Sciences (20%) and the Humanities (20%), than in Life Sciences (16%) and Technological Sciences (4%) (Ferreras-Fernández, García-Peñalvo, & Merlo-Vega, 2015). Finally, an analysis of 612 digitised engineering theses and dissertations at the North-West University in South Africa during 2002 to 2014 showed that about 41% had at least one Google Scholar citation (Bangani, 2018).

Despite the above investigations, no study has shown that it is possible to systematically count the citations to large numbers of theses from a source that has reasonable coverage of non-article scholarly documents, such as Google Scholar, and nothing is known about the types of document that cite theses for academia in general.

## 2.3  Downloads or views of electronic theses

Statistics about downloads or views have been used to assess the usage of electronic dissertations (e.g., Zhang, Lee, & You, 2001; Coates, 2013; Ferreras-Fernández, García-Peñalvo, & Merlo-Vega, 2015). For instance, the University College London repository gives detailed downloaded statistics for dissertations[8] and the ProQuest Dissertations Dashboard service (https://dissdash.proquest.com) allows universities providing dissertation data to access usage statistics (Patel et al., 2015). One study compared the download count and print circulation of the digitised dissertations at the University of Massachusetts Amherst, finding that the average download count of dissertations for the past two years was almost 13 times higher than the average print circulation of dissertations for the last 10 years (Adamick, 2016). The average PDF views of 621 engineering doctoral dissertations was found to be 215 times much higher than their average Google Scholar citation counts (Bangani, 2018), suggesting that statistics of views or full-text downloads may reflect wider usage of a published scholarly work, such as by students, some non-citing academics, and even non-academic users (Kurtz & Bollen, 2010).

## 2.4  Mendeley readership counts

The social reference sharing site Mendeley (Henning & Reichelt, 2008) provides a convenient source of readership data (e.g., Kudlow, Cockerill, Toccalino, Dziadyk, Rutledge, Shachak, & Eysenbach, 2017; Zahedi, & van Eck, 2018) from the minority of researchers that use it (Van Noorden, 2014). Many studies have shown that Mendeley readership counts positively correlate with counts of citations to published journal articles (e.g., Costas, Zahedi, & Wouters, 2015; Thelwall, Haustein, Larivière, & Sugimoto, 2013; Zahedi, Costas, & Wouters, 2014). There is also evidence that early Mendeley reader counts correlate with longer term citation counts (e.g., Thelwall & Sud, 2016; Thelwall, 2018) and that users can register types of document that they do not necessarily cite, such as editorials (Zahedi & Haustein, 2018). Hence, Mendeley readership count might be useful to assess the early impact of doctoral dissertations, partly avoiding the publication delays of citation counts. Moreover, many authors or students may add dissertations to their Mendeley libraries when reading them for non-citing reasons, such as teaching and learning, which might be useful to reflect educational uses of doctoral dissertations (Mohammadi, Thelwall, & Kousha, 2016). It seems that no study has investigated Mendeley readers of dissertation so far, although they have been shown to have low or moderate correlations with Google Books citations to academic monographs in many fields (Kousha & Thelwall, 2016).

# 3  Research questions

The research goal is to assess the value of Google Scholar for dissertation citation analysis based on its new partnership with ProQuest. In addition to the basic ability to find citations, the types of citations are important to interpret the nature of their impact and the types of people that read them, especially those that do not cite. The following questions drive address this issue.

---

8 . http://discovery.ucl.ac.uk/10057177/

1. Can citations to ProQuest doctoral dissertations be semi-automatically extracted from Google Scholar with a high degree of accuracy?
2. Is Google Scholar better than Mendeley for dissertation scholarly impact calculations?
3. Do Google Scholar citations and Mendeley readers reflect a similar type of impact for doctoral dissertations?
4. How accurately does Google Scholar identify citations to dissertations and which types of scholarly document cite dissertations?
5. Which types of people read dissertations?

# 4  Methods

This section introduces a practical method to systematically identify Google Scholar citations to dissertations for large scale evaluations from *the ProQuest Dissertation & Theses* database. Because Google Scholar does not support automatic API searches, a new method is introduced using the *Publish or Perish* software (Harzing & van der Wal, 2008) with pre-defined queries to limit searches to ProQuest dissertations in a manageable manner.  All data collection was conducted during November 2018.

## 4.1  Google Scholar queries for ProQuest dissertations

To search Google Scholar for citations to ProQuest doctoral dissertations a combination of the *site:* search command, phrase searches, and the *author:* search command was used and the results were limited by year, as described below.

The **site:proquest.com** command was used as the basic Google Scholar query to limit the results to the ProQuest website. Since the site:proquest.com command in Google Scholar retrieves many non-dissertation publications from ProQuest databases (e.g., journal articles and conference papers), it was necessary to refine the query to eliminate these.

The phrase search **"*The quality of this reproduction is dependent upon*"** was added to the *site:proquest.com* command because this copyright information is found in most ProQuest dissertations (see Figure 1). To check how common the above text was, this text was searched as a phrase in the main ProQuest Dissertations & Theses database through an institutional subscription. Of 12,518 full-text English language doctoral dissertations published during 2013-2017, 91% (11,339) included the above phrase, suggesting that the phrase search gives high coverage. Only part of the copyright statement was used in the queries because Google Scholar has a 256 character query limit (Boeker, Vach, & Motschall, 2013) and extra information also had to be added to the query (see below).

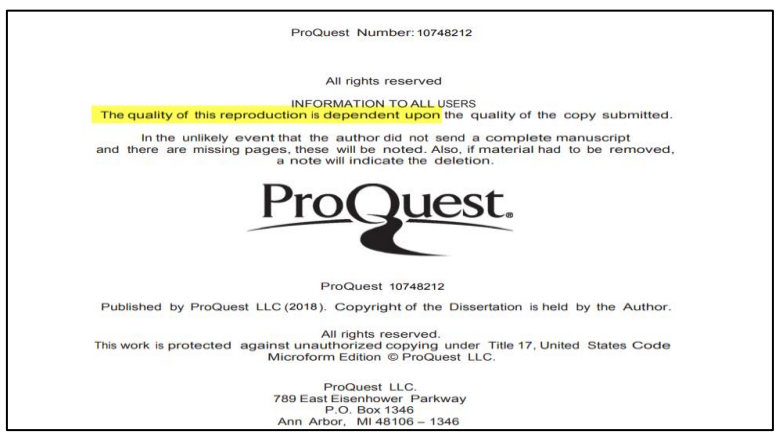

Figure 1. An example of a copyright page in ProQuest dissertations illustrating the query phrase used (yellow-highlighted) to identify ProQuest dissertations via Google Scholar searches

To limit the results to doctoral dissertations, the phrase search, "**Doctor of**", was added to the query (i.e., *site:proquest.com "The quality of this reproduction is dependent upon" "Doctor of"*).  This phrase was selected instead of "*Doctor of Philosophy*" to cover a range of doctorates, as shown in the examples below:

- *Submitted in partial fulfillment of the requirements for the degree of **Doctor of** Philosophy*
- *Submitted in partial satisfaction of the requirements for the degree of **Doctor of** Nursing*
- *Submitted for the degree of **Doctor of** Medicine*
- *Submitted for the award of **Doctor of** Education*
- *Submitted for the degree of **Doctor of** Psychology*
- *Prepared for the Degree of **Doctor of** Musical Arts*
- *Submitted in candidacy for the Degree for **Doctor of** Business Administration*

This technique allows some false matches. For instance, it may retrieve master's theses in medicine where the term "Doctor of" is mentioned in the text in other contexts. Hence, it was necessary to match the data from Google Scholar against the original ProQuest Dissertations & Theses database to exclude false matches.

Google Scholar only shows the first 1,000 results, and so searches were limited to different years using the Google Scholar Advanced Search option. Nevertheless, this method was not always enough to reduce the Google Scholar results below 1,000. For instance, the query *site:proquest.com "The quality of this reproduction is dependent upon" "Doctor of"* returned over 16,000 matches for the year 2013. Hence, an extra method was needed to extract all Google Scholar citations to ProQuest dissertations for each year, as follows.

A query splitting method was used to restrict the Google Scholar ProQuest dissertation search results to below 1,000, by adding author first name initials (A-Z) to the ***author:*** command. This matches the letter anywhere in author first names or initials. For instance, the query *site:proquest.com "Doctor of" "The quality of this reproduction is dependent upon" author:Z* locates ProQuest dissertations from Google Scholar where the letter **Z** is anywhere in initials of authors' first names, such as "**Z** Lai", "**Z**B Haber",  "J**Z** Fuller" or "M**Z**A Durrani" as shown in Figure 2.

Figure 2. A query splitting method using the *author:Z* command to extract Google Scholar citations to doctoral ProQuest dissertations with an author first name containing 'Z'.

Some letters were more frequently used in the initials of authors' first names, returning more than 1,000 hits for each year. Hence, twenty six manual Google Scholar searches were conducted using the author:[A to Z] command in the pre-defined query to estimate total number of Google Scholar hits for ProQuest doctoral dissertations in 2013. Table 1 shows that 12 letters returned more than 1,000 hits in Google Scholar. Letters with accent marks (e.g., Á, Ä É, Ë, Ç and Ö) mostly returned no results, with minor exceptions, such as *author:Á* with two hits during 2013-2017. It seems that Google Scholar usually ignores letters with accent marks (see also https://harzing.com/resources/publish-or-perish/manual/about/faq).

Table 1. The frequency of different letters in the initials of authors' first names using the Google Scholar citation queries for ProQuest doctoral dissertations published in 2013*.

| Letters used in a*uthor:* search command | Google Scholar hits | Letters used in a*uthor:* search command | Google Scholar hits |
|---|---|---|---|
| **M** | **3,130** | H | 808 |
| **A** | **2,880** | G | 739 |
| **J** | **2,880** | N | 721 |
| **S** | **2,250** | W | 657 |
| **C** | **1,850** | F | 512 |
| **L** | **1,800** | Y | 485 |
| **D** | **1,710** | V | 372 |
| **R** | **1,600** | O | 252 |
| **K** | **1,420** | I | 242 |
| **E** | **1,370** | X | 159 |
| **B** | **1,070** | Z | 136 |
| **T** | **1,040** | Q | 89 |
| P | 961 | U | 30 |

\* *site:proquest.com "Doctor of" "The quality of this reproduction is dependent upon" author:[A-Z]*

To resolve this issue, the Google Scholar results were sorted from highest to lowest for each year (e.g., Table 1 for 2013) and then searched using the letters giving the fewest Google Scholar hits via the query described above: *site:proquest.com "Doctor of" "The quality of this reproduction is dependent upon"* ***author:U***. Then, *-author:* command was added to subsequent queries to exclude any previously searched letters, as in the following example: *site:proquest.com "Doctor of" "The quality of this reproduction is dependent upon" author:Q* ***-author:U***. This technique reduced the total number of hits to below 1,000 for each query, enabling a complete set of results to be extracted. For instance, the method reduced the number of Google Scholar dissertation matches for the author first name initial **M** from 3,130 (see Table 1) to 765. Because Google Scholar has a 256-character query limit, for some letters with the most Google Scholar hits (e.g., M, A, J or S) it was not possible to exclude all previously searched letters. For instance, for the query below to retrieve ProQuest dissertations with M in the initials of the authors' first names, only 16 other letters A-W were excluded before reaching the 251-character query limit (just 5 characters less than maximum 256 query limit). Therefore, the query splitting stage produced some duplicate Google Scholar records and it was necessary to delete identical titles in the final stage.

*site:proquest.com "Doctor of" "The quality of this reproduction is dependent upon"* ***author:M*** *-author:A -author:J -author:S -author:C -author:L -author:D -author:R -author:K -author:E -author:B -author:T -author:P -author:H -author:G -author:N -author:W*

## 4.2   Automatic Google Scholar dissertation searches

*Publish or Perish* retrieves and analyzes academic citations from Google Scholar (Harzing, 2007; Harzing & van der Wal, 2008). The above procedure was used to generate queries that were submitted by *Publish or Perish* for each letter and year separately. To avoid problems with sending too many requests to Google Scholar, the data collection process was spread over a week with the recommended delays between queries (https://harzing.com/resources/publish-or-perish/manual/about/faq).

## 4.3   Data cleaning and subject matching

Dissertations extracted from Google Scholar were matched with the *ProQuest Dissertations & Theses* database to get subject, degree, and country information. For this, queries were submitted in the Document Title field ("TI") of the ProQuest "Command Line Search" option as phrase searches based on the titles of the dissertations gathered from Google Scholar searches. Because the maximum number of Boolean operators allowed in a single ProQuest search was 1,000, eighty-four separate searches were conducted and the combined results covered all 83,926 titles from the Google Scholar dataset, as in the example below.

> **TI**("identifying effective education interventions in sub saharan Africa a meta analysis of rigorous impact evaluations") OR **TI**("precast column footing connections for accelerated bridge construction in seismic zones") OR **TI**("dual fuel reactivity controlled compression ignition rcci with alternative fuels") OR …

Of 83,926 records extracted from the Google Scholar dissertation searches, 97% (81,533) were matched with data from the original ProQuest Dissertations & Theses database based on their titles and the last names of their authors. Because the phrase search "Doctor of" to retrieve only doctoral dissertations from Google Scholar searches can also retrieve false matches (Master's dissertations), the field "Degree" from the main ProQuest dissertation data was used to filter out the 1.4% (1,111 of 81,533) non-doctoral theses such as M.A. (Master of Arts), M.S. (Master of Science), M.P.H. (Master of Public Health) and M.P.P. (Master of Public Policy). Nevertheless, the method used here had a high overall level of accuracy (98%) for locating doctoral ProQuest dissertations from Google Scholar searches without the final filtering stated. From the 80,422 remaining doctoral dissertations in the dataset, 97% (77,884) were from the United States and 3% (2,538) were from other countries (1% from UK, 0.8% from Singapore, 0.5% from Canada, 0.2% from Australia and 0.5% from other countries), indicating that almost all ProQuest dissertations indexed by Google Scholar were from American universities. One reason is that ProQuest largely covers dissertations from American universities. Another reason might be that at the time of data collection (November 2018) not all the doctoral dissertations provided by ProQuest were indexed by Google Scholar (see Arbor, 2017). To have a more uniform dataset for statistical analysis, the 77,884 dissertations from American universities were analysed and the rest were discarded.

The Subject field of the main ProQuest Dissertations & Theses database was used for disciplinary analyses[9]. However, because there were too few dissertations per subject and year, the OECD classification scheme[10] was used to re-categorize the ProQuest doctoral dissertations into 18 broad subjects. For instance, all relevant engineering fields (e.g., *Engineering, Civil engineering, Mechanical engineering, and Electrical engineering)* in ProQuest output were combined to form the broad subject "*Engineering and Technology*". Similarly, all related Art subjects (e.g., *Music, Architecture, Performing Arts, Dance and Theater*), Medical Sciences (e.g., *Medicine, Pharmacy sciences, Surgery, Toxicology, Immunology, and Microbiology*) and Social Sciences (e.g., *Political science, Social research Sociology, Communication, and Information science*) were combined into broad subject areas. Appendix A shows the number of American ProQuest doctoral dissertations identified by Google Scholar from 2013-2017 in 18 fields.

## 4.4   Mendeley reader counts

Mendeley readership counts were used to estimate the wider or early impact of dissertations in addition to the Google Scholar citations. All 77,884 dissertations were submitted to Mendeley via its API in the free software *Webometric Analyst* (http://lexiurl.wlv.ac.uk) with the record the total Mendeley reader count. Initial tests revealed that the most appropriate method to identify matching dissertation records in Mendeley is through metadata searches (author names and title). Therefore, each dissertation was queried by its title and author last name as shown in the example below:

> **title**:Correlational study of risk management and information technology project success AND **author**:Gillespie

A set of filtering rules was used to filter out false Mendeley reader counts. For instance, the educational sciences dissertation, "*Teachers' perceptions and implementation of professional learning communities in a large suburban high school*" by Peppers, G. had 15 Mendeley readers, based on an article published in *National Teacher Education Journal* with the same title and author. The "Source" or "Type" fields in Mendeley output were therefore used to limit the results to dissertations (e.g., "Thesis", "PhD Thesis", "ProQuest Dissertations and Theses", "Doctoral Dissertation", "Dissertation", "Dissertation Abstracts International", or "PQDT", "PhD Thesis, Columbia University") with manual checks. Of 13,247 Google Scholar records with at least one Mendeley reader count, 95.4% (12,632) were identified as dissertations and the remaining 4.6% (615) were discarded.

# 5   Results

As the methods section shows, large numbers of doctoral dissertations and their citation counts can be extracted semi-automatically from Google Scholar using curated queries and with the aid of Publish or

---

Perish. The above method only located 30% (77,884 of 264,149) of the American doctoral dissertations indexed in ProQuest Dissertations & Theses for 2013-2017, however. It is not clear whether this is because Google Scholar has not yet or will never index all of ProQuest or because the above method has low recall. This issue is addressed in the Discussion section.

Overall, about a fifth of the ProQuest doctoral dissertations from American universities during 2013-2017 (n=77,884) had at least one Google Scholar citation (20%: 15,860) or Mendeley reader (16%: 12,632). Higher proportions of dissertations published during 2013-2015 had been cited in Google Scholar than had Mendeley readers (Figure 3). This suggests that Google Scholar citations could be a more useful indicator for dissertations at least 3-5 years after publication. In contrast, Mendeley readership counts seem to be more useful for more recent dissertations (1-2 years after publication), before they had time to be cited much by other publications (Figure 1). This early impact advantage for Mendeley supports previous studies of journal articles (Thelwall & Sud, 2016; Maflahi & Thelwall, 2017; Thelwall, 2018). To investigate disciplinary differences, both the average number (geometric mean) and proportion of Google Scholar citations and Mendeley readers were calculated for each fields and years (see below).

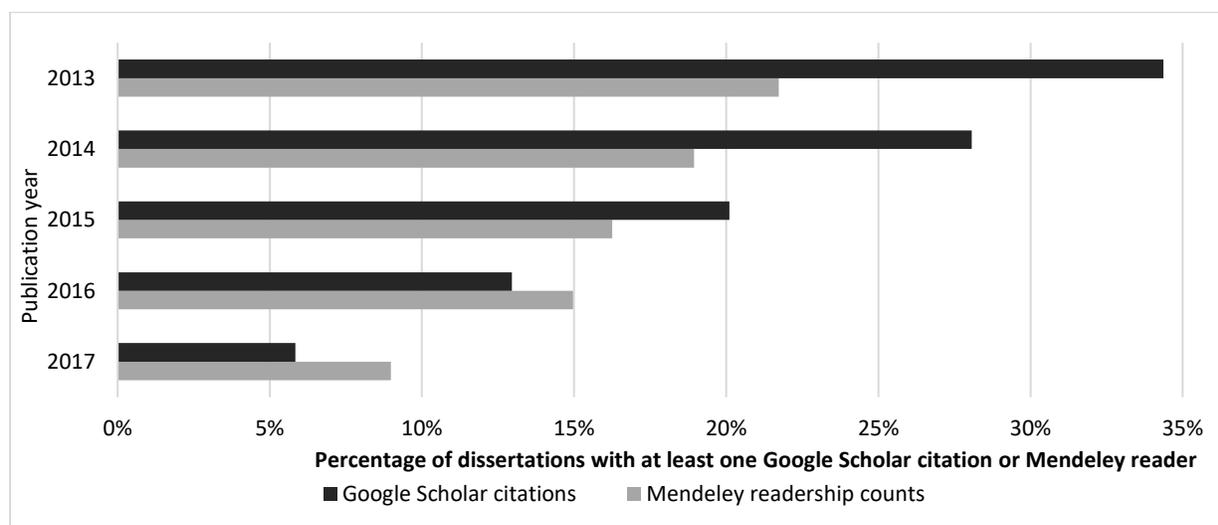

Figure 3. Proportions of ProQuest doctoral dissertations published during 2013-2017 from American universities with at least one Google Scholar citation and Mendeley reader.

## 5.1 Average Google Scholar citation counts and Mendeley reader counts

Geometric means were used to compare the average number of Google Scholar citations and Mendeley readers for the ProQuest dissertations (Figures 4-8). Geometric means were used instead of the arithmetic mean and median because it is a better central tendency indictor for highly skewed data, distinguishing differences more effectively within datasets with many zeros. The geometric mean was calculated by 1) adding 1 to the citation and Mendeley reader counts $c_i$, 2) taking the natural log, 3) calculating the arithmetic mean of this transformed data, 4) applying the exponential function to the result, and 5) subtracting 1 (Thelwall, 2016a). This procedure $\exp\left(\overline{ln(c_i + 1)}\right) - 1$ was repeated for each field and year separately. Confidence intervals were calculated at the 95% level to suggest when differences between average Google Scholar citation counts and average Mendeley readership counts are statistically significant. The confidence intervals used the standard normal distribution formula $\overline{ln(c_i + 1)} \pm$ $t_{n-1} \, s/\sqrt{n}$ where $n$ is the sample size (number of articles), $t_{n-1}$ is the critical value of the t distribution at the 0.05/2 level, $s$ is the standard deviation of the log-transformed data $ln(c_i + 1)$, and the lower and upper limits $l$ were converted by applying the formula $exp(l) - 1$. The graphs show that there are enormous disciplinary differences in the average citation counts of dissertations (based on all the document types indexed by Google Scholar).

In the eight social sciences, arts and humanities fields the geometric mean number of Google Scholar citations to doctoral dissertations published in 2013, 2014 and 2015 (with 3-5 year time window for citation analyses) were 1.7, 1.6 and 1.4 times higher than in the 10 science, technology and biomedical subjects respectively. Moreover, the average numbers of Google Scholar citations to 2013, 2014 and 2015 dissertations were highest among all science and biomedical fields in Engineering and Technology (0.52; 0.41; 0.30) and Earth and Environmental Sciences (0.46; 0.45; 0.27) and lowest in Medical Sciences (0.17; 0.13; 0.07), Biological Sciences (0.14; 0.11; 0.09) or Chemical Sciences (0.13; 0.11; 0.08). Hence, it seems that doctoral dissertations tend to be more cited in most social sciences, arts and humanities subjects, in addition to Engineering and Environmental Science, in comparison to other fields. Google Scholar could be useful alternative platform to capture research impact in these areas.

On average, there are more Google Scholar citations than Mendeley readers in most fields for older dissertations published 2013-2014 (Figures 4, 5). For instance, the geometric mean numbers of Google Scholar citations are clearly higher than Mendeley reader counts in 11 out of 18 fields for dissertations published during 2013 and 2014. In contrast, the average Mendeley reader counts are higher than the average Google Scholar citation counts for recently published dissertations in 2017. However, the geometric mean number of Mendeley readers is also higher than the Google Scholar citation count in 7 fields for dissertations in 2017 (Figure 8). This is probably because dissertations need time (at least three years) to be cited by other publications while Mendeley readers can be recorded without publication delays. There are also clear disciplinary differences. For example, Mendeley had an advantage over Google Scholar across all years by finding more readers of doctoral dissertations in Medical Sciences, Health Sciences, Psychology and Social Sciences. In Engineering and Technology and Mathematics fields Google Scholar also found more citations than Mendeley readers across all years.

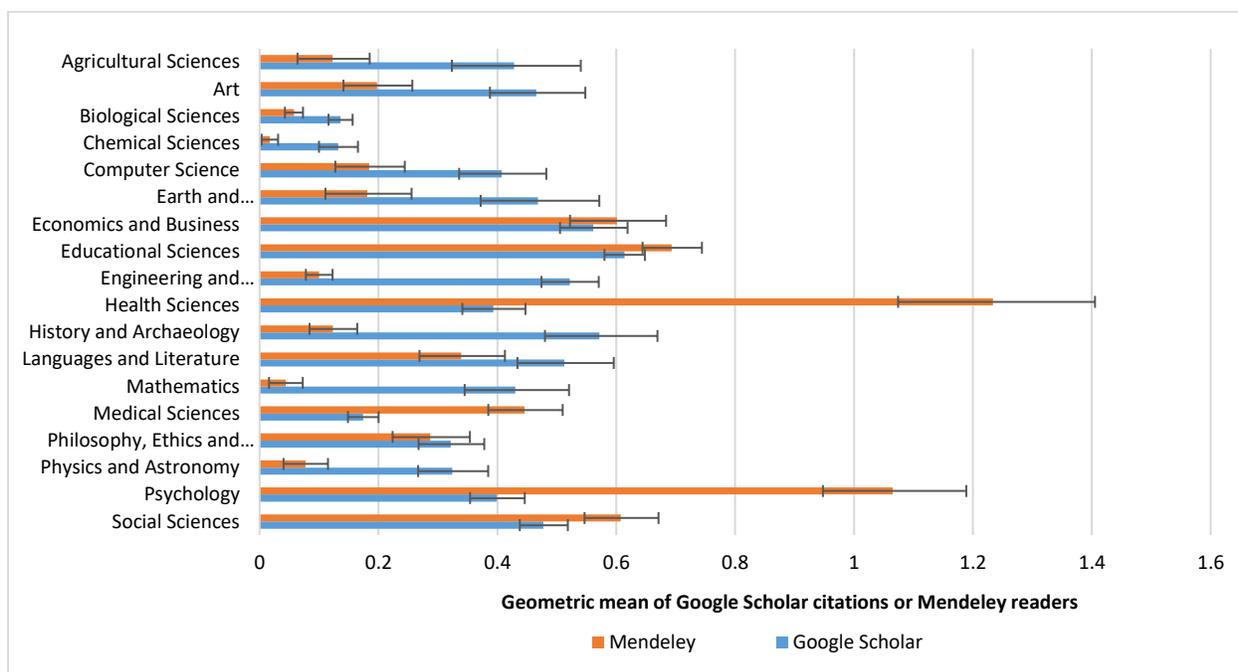

Figure 4. Geometric mean number of Google Scholar citations and Mendeley readers for ProQuest doctoral dissertations in 2013 across 18 fields.

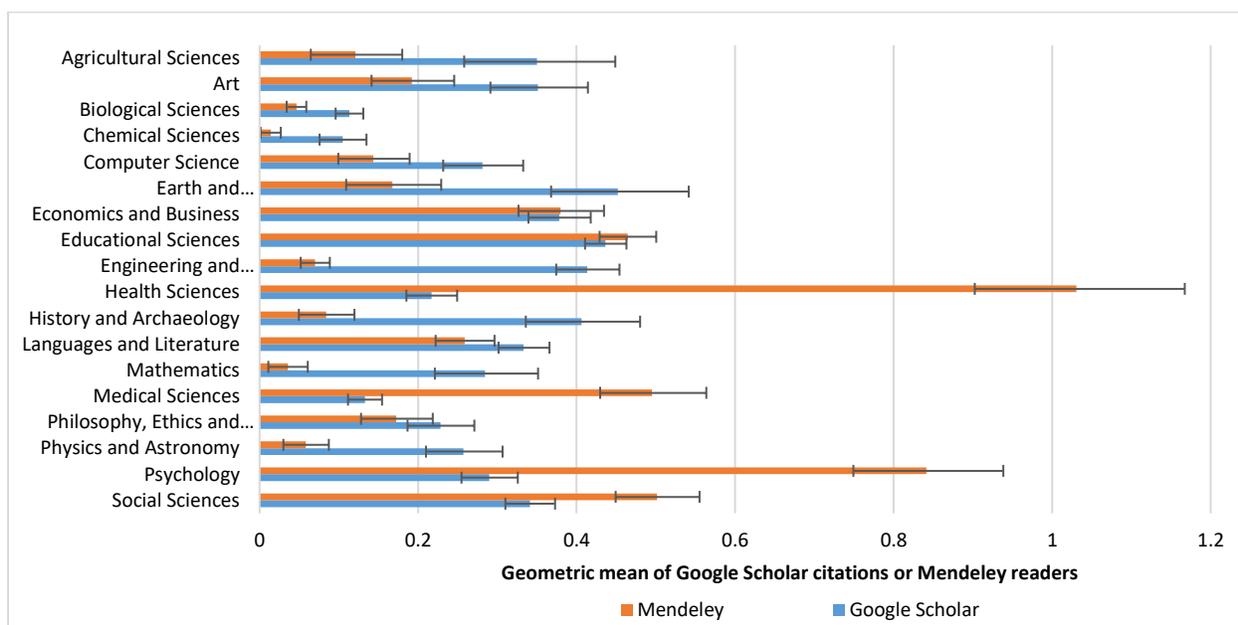

Figure 5. Geometric mean number of Google Scholar citations and Mendeley readers for ProQuest doctoral dissertations in 2014 across 18 fields.

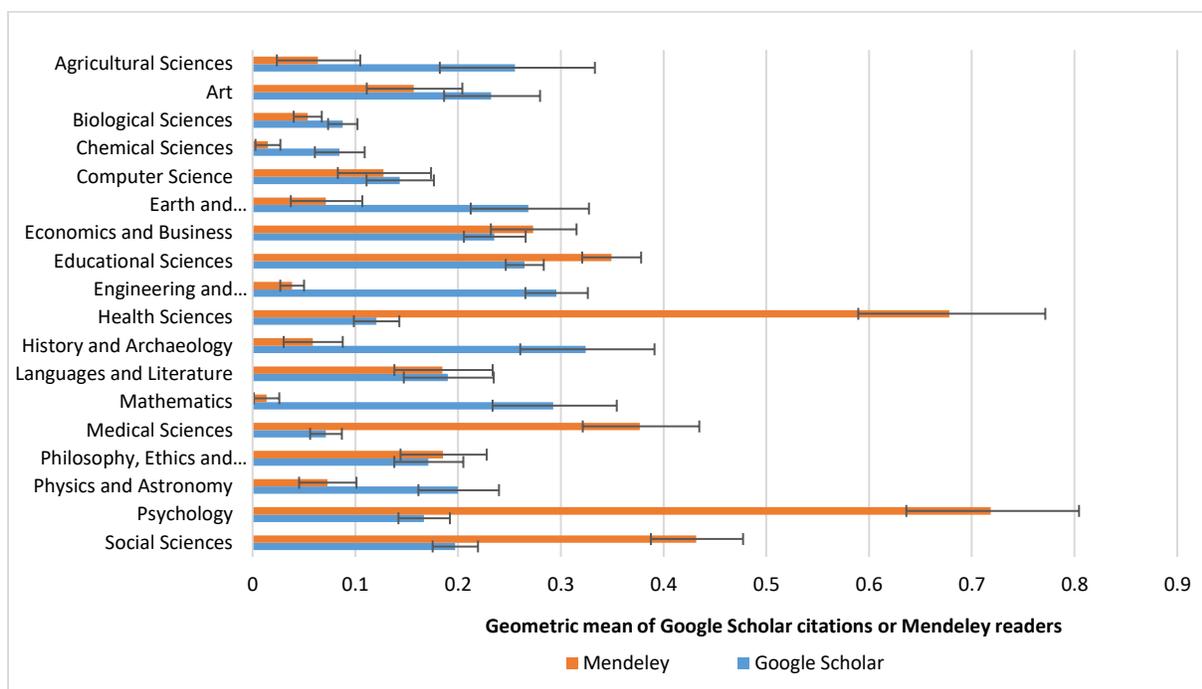

Figure 6. Geometric mean number of Google Scholar citations and Mendeley readers for ProQuest doctoral dissertations in 2015 across 18 fields.

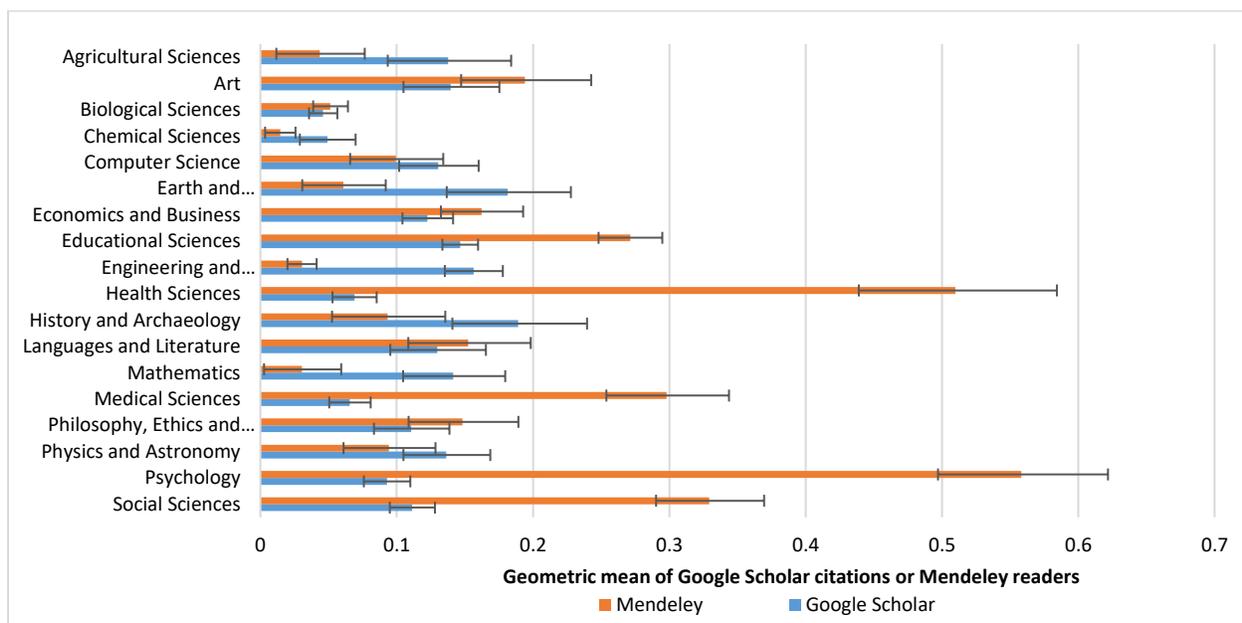

Figure 7. Geometric mean number of Google Scholar citations and Mendeley readers for ProQuest doctoral dissertations in 2016 across 18 fields.

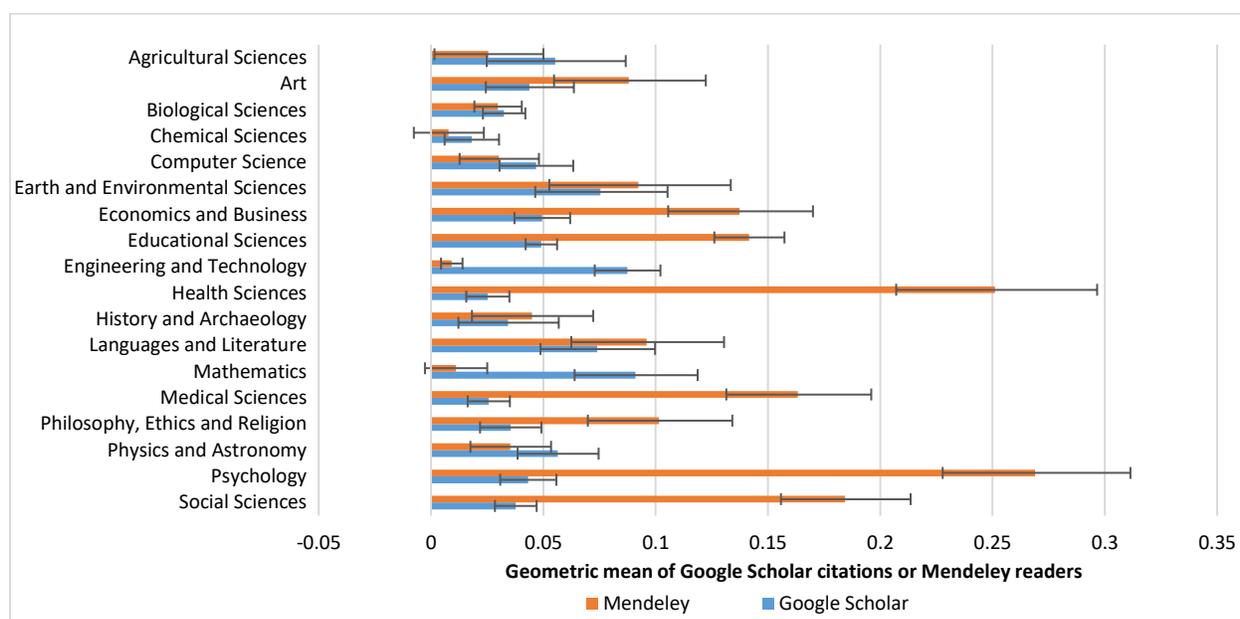

Figure 8. Geometric mean number of Google Scholar citations and Mendeley readers for ProQuest doctoral dissertations in 2017 across 18 fields.

## 5.2 Proportions of dissertations with Google Scholar citations or Mendeley readers

In general, the proportions of doctoral dissertations with at least one Google Scholar citation in the eight social science, arts and humanities fields during 2013, 2014 and 2015 were 1.6, 1.5 and 1.4 times higher than in the 10 science, technology and biomedical disciplines, respectively (Figures 9-11). However, probably because most dissertations published within 2016-2017 had less time to be cited by other publications, the proportion of Google Scholar cited dissertations is low in recent years (Figures 12-13).

The proportion of Google Scholar citations to dissertations in 2013, 2014, 2015 and 2016 were statistically higher (in the sense of non-overlapping confidence intervals, even though this is not the same as a hypothesis test) than the proportion with Mendeley readers in 15, 14, 12 and 8 subject areas out of the 18 fields respectively, whereas the proportion with Mendeley readers was only higher than the proportion with Google Scholar citations for 2014 in two fields (Medical Sciences and Health Sciences), for 2015 in three fields (Medical Sciences, Health Sciences, and Psychology) and for 2016 in four fields (Medical Sciences, Health Sciences, Psychology and Social Sciences) at the 95% confidence limits. In contrast, Mendeley had a bigger advantage over Google Scholar for more recently published dissertations from 2017, finding more doctoral dissertations with at least one reader in 6 out of 18 fields (Figure 13). These results suggest that Google Scholar citations could be useful for impact assessment of older dissertations and Mendeley may help as early evidence of readership impact, when they do not have enough time to be cited by academic publications.

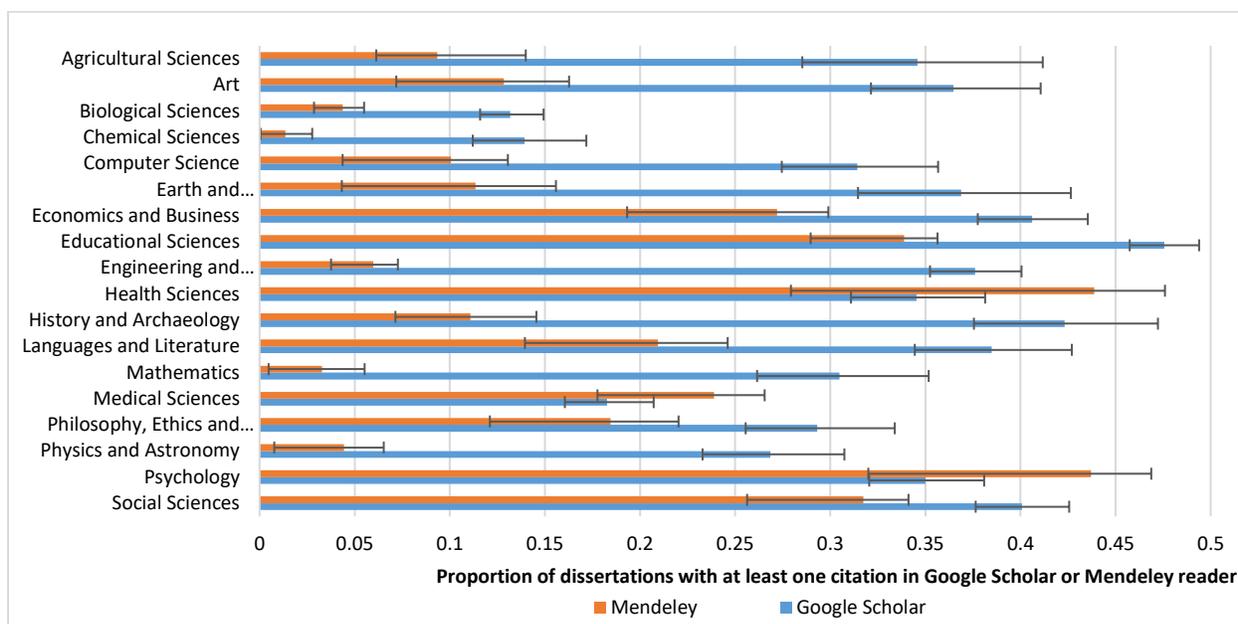

Figure 9. Proportion of ProQuest doctoral dissertations in 2013 with at least one citation in Google Scholar or Mendeley reader across 18 fields.

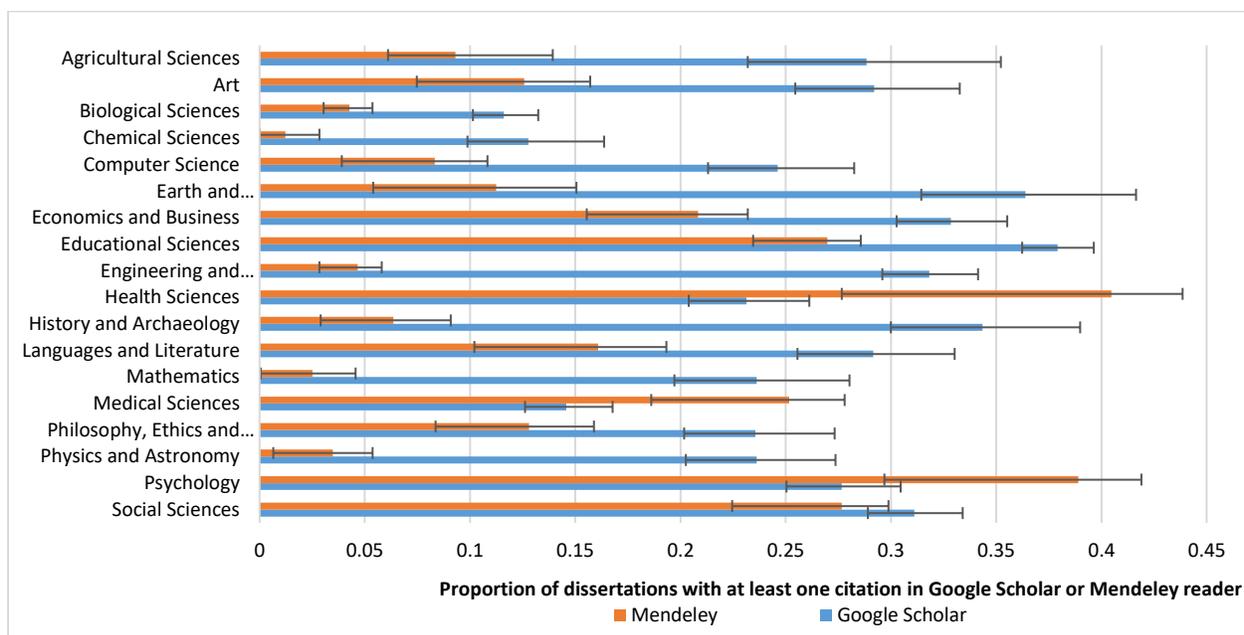

Figure 10. Proportion of ProQuest doctoral dissertations in 2014 with at least one citation in Google Scholar or Mendeley reader across 18 fields.

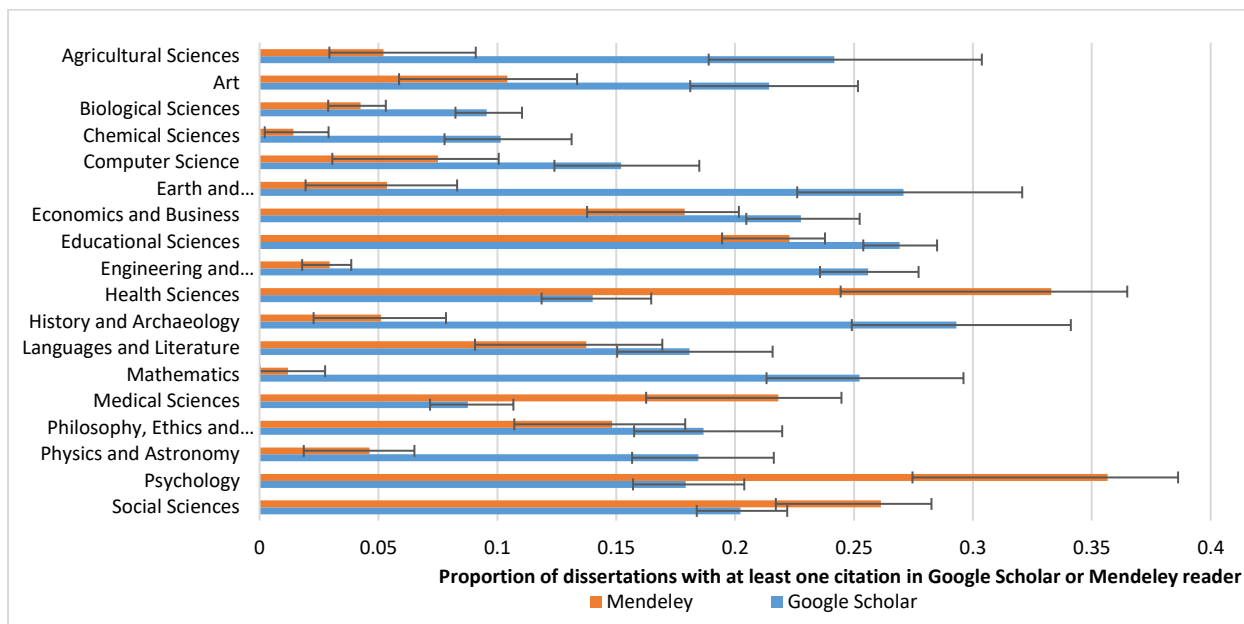

Figure 11. Proportion of ProQuest doctoral dissertations in 2015 with at least one citation in Google Scholar or Mendeley reader across 18 fields.

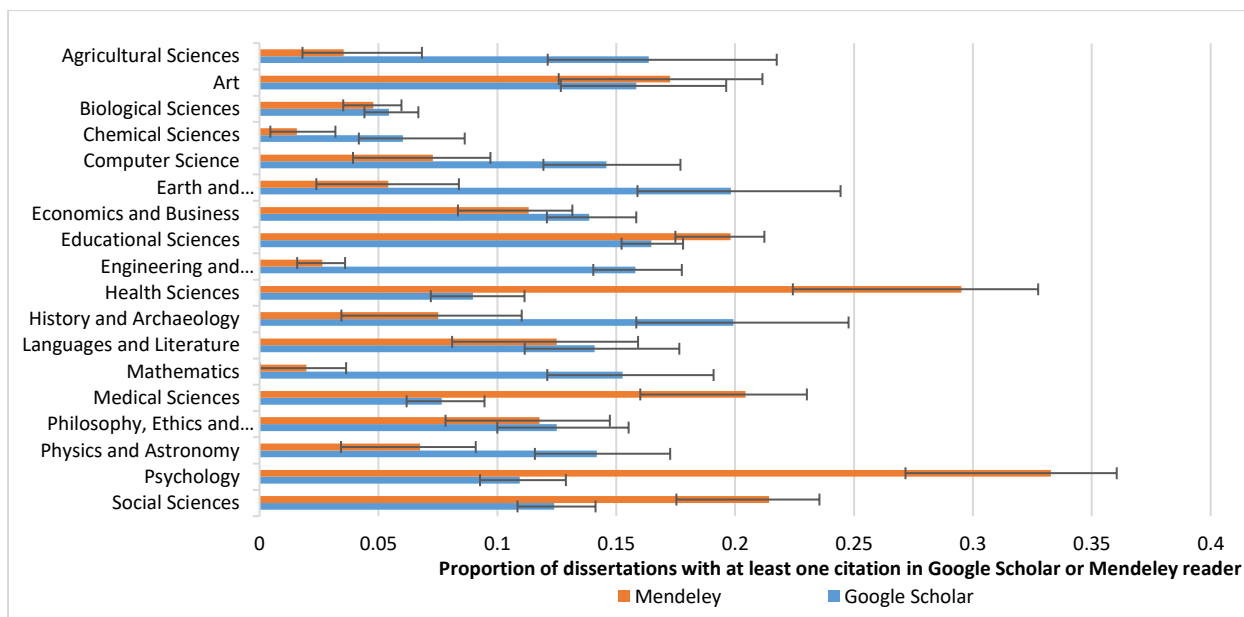

Figure 12. Proportion of ProQuest doctoral dissertations in 2016 with at least one citation in Google Scholar or Mendeley reader across 18 fields.

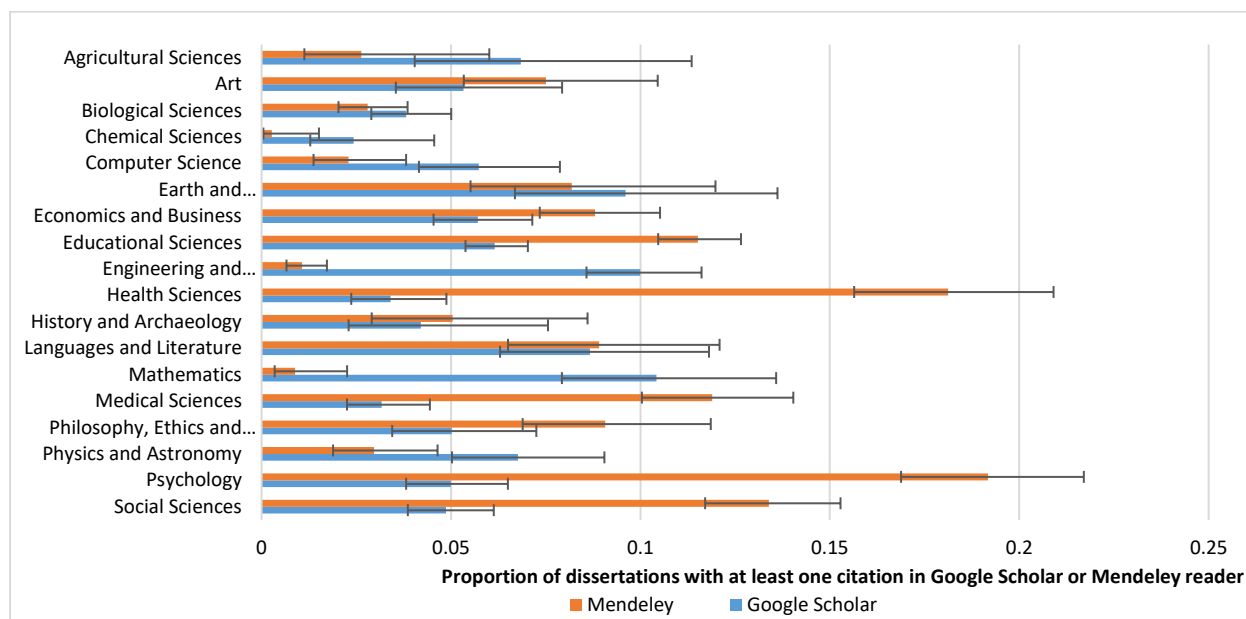

Figure 13. Proportion of ProQuest doctoral dissertations in 2017 with at least one citation in Google Scholar or Mendeley reader across 18 fields.

## 5.3   Correlations between Google Scholar citations and Mendeley readers

Assessing the correlations between Google Scholar citations and Mendeley readers for dissertations can provide statistical evidence about whether they may reflect similar types of impact. Spearman correlation tests were used instead of Pearson because the frequency distributions of both Google Scholar citation counts and Mendeley reader counts were highly skewed. Correlations were calculated separately for 18 subject areas and each year during 2013-2017 to discount the impact of time and disciplinary differences on the magnitude of correlations (Fairclough & Thelwall, 2015).

In general, there were low or insignificant correlations between Google Scholar citation counts and Mendeley reader counts for doctoral dissertations in most fields and years (Table 2), suggesting that the two indicators only loosely reflect similar types of impact. This result is in contrast with previous studies that have found citation counts and Mendeley reader counts for journal articles to have moderate or strong associations in most fields, except for young articles (e.g., Haustein, Larivière, Thelwall, Amyot, & Peters, 2014; Mohammadi & Thelwall, 2014; Zahedi, Costas, & Wouters, 2017). The magnitude of the correlations may partly reflect the relative scarcity of both readers and citations, since low numbers reduce correlations in discrete data sets (Thelwall, 2016b). The low numbers are presumably due to associated articles being frequently cited rather than dissertations. For example, the dissertation, "Population Health of California Fishers" by Mourad Gabriel from 2013 has 7 Google Scholar citations but a co-authored paper by the candidate from 2012, with the same title as Chapter 3 of the dissertation, "Anticoagulant Rodenticides on our Public and Community Lands: Spatial Distribution of Exposure and Poisoning of a Rare Forest Carnivore", has 97 Google Scholar citations.

Of the 20 dissertations with the most Google Scholar citations (citation range 38-20), 13 had no Mendeley readers. For instance, the Ph.D. dissertation, *Trajectory design and orbit maintenance strategies in multi-body dynamical regimes* by Thomas Pavlak published in 2013 had 33 Google Scholar citations without any Mendeley readers. Conversely, the doctoral thesis, "*Evaluation of a telehealth intervention combining structured self-monitoring of blood glucose and nurse care coordination among people with type 2 diabetes noninsulin-treated*" by Deborah Greenwood published in 2014 had 51 Mendeley readers but no Google Scholar citations. Students or academics may add relevant doctoral dissertations to their Mendeley library for non-citation reasons, such as for teaching, course assignments or professional activities (see Mohammadi, Thelwall, & Kousha, 2016). Nevertheless, there are significant correlations, albeit low, between Google Scholar citations and Mendeley readers in 11 out of 18 fields for dissertations published in 2013, suggesting that both indicators may better reflect similar types of intellectual impact for older dissertations. Stronger correlations are more common in Computer Science and Educational Sciences during 2013-2016 and in Economics and Business, Languages and Literature, and Medical Sciences during 2013-2015, suggesting disciplinary differences in the association between Google Scholar citations and Mendeley readers.

Table 2. Spearman correlations between Google Scholar citation counts and Mendeley reader counts for ProQuest doctoral dissertations by year and field.

| Subject area | 2013 | 2014 | 2015 | 2016 | 2017 |
|---|---|---|---|---|---|
| Agricultural Sciences | .103** | .043 | .016 | -.024 | .237 |
| Art | .079 | .061 | .126** | .09 | .018 |
| Biological Sciences | .051* | .027 | .002 | -.027 | .040 |
| Chemical Sciences | .099* | .089 | .079 | -.032 | .031 |
| Computer Science | .117** | .128** | .090* | .089* | .008 |
| Earth and Environmental Sciences | -.024 | .005 | -.03 | -.019 | .081 |
| Economics and Business | .143** | .145** | .146** | -.005 | .019 |
| Educational Sciences | .077** | .061** | .044* | .046* | .025 |
| Engineering and Technology | .145** | .076** | -.009 | .031 | .031 |
| Health Sciences | .049 | .038 | .133** | .065 | .017 |
| History and Archaeology | -.051 | .055 | -.036 | .100 | .049 |
| Languages and Literature | .099* | .116** | .097* | -.013 | -.062 |
| Mathematics | .050 | .083 | .034 | .087 | -.032 |
| Medical Sciences | .117** | .164** | .140** | .053 | -.013 |
| Philosophy, Ethics and Religion | .059 | -.004 | .044 | .062 | .052 |
| Physics and Astronomy | .039 | -.014 | .070 | .048 | -.010 |
| Psychology | .083* | .038 | .012 | .019 | .037 |
| Social Sciences | .069** | .050* | .031 | .025 | .019 |

\* Significant at the p = .05 level. \*\* Significant at the p = .01 level.

## 5.4 Characteristics of documents indexed by Google Scholar that cite dissertations

The citations to ProQuest dissertations found by Google Scholar were manually checked to estimate 1) the accuracy of Google Scholar for identifying citations to dissertations in the references of other publications, 2) citing source type (journal article, conference paper, book/book chapter, thesis, or other) and 3) the extent of self-citation. For this, a sample of 20 dissertations with at least one Google Scholar citation were selected from each of the 18 fields during 2013-2017 (n=360 dissertations). This sample was used for an overall qualitative analysis of the Google Scholar citations to dissertations in a manageable manner (see Discussion). Since not all fields have the same number of ProQuest dissertations, results from it are representative of a field average rather than an overall average (see Appendix A). The 360 sampled articles were cited by 704 publications, as reported by Google Scholar. For 646 (92%) of these, the full text could be accessed either through an institutional subscription or an open access copy or the references could be found through the cited reference list in Scopus. Manual checks of these citing sources found that Google Scholar identified citations to doctoral dissertations with a high level of precision: 96.6% (624 out of 646). Only 22 errors were found in Google Scholar. For instance, two of the three Google Scholar citations to the dissertation, "*Guaranteed adaptive univariate function approximation*" did not exist: the dissertation was not mentioned anywhere in the text. A common Google Scholar error was reporting a citation to a dissertation when there was a citation to a journal article or conference paper with a similar title and authors (e.g., for the dissertation: "*Widely tunable Sampled Grating Distributed Bragg Reflector Quantum Cascade laser for gas spectroscopy applications*"). The estimated overall level of precision is 97.0%, after correcting for the samples of 20 not being proportional to the number of dissertations in each field.

The manual checks found that 22% of the citations to dissertations were self-citations from other publications that were authored or co-authored by the dissertation author. Correct citations originated from journals (56%), dissertations (29%), books and monographs (6%), conference papers (5%) and other publications (4%). However, in eight social science and the arts and humanities fields, there was a higher proportion of citations to dissertations from other dissertations (33%) and books (10%) than in the 10 science subjects (26% from dissertations and 3% from books).

## 5.5 Readership status of doctoral dissertations in Mendeley

Mendeley reports the self-declared occupations of readers, which can give useful extra information about the impact of dissertations. The Mendeley reader proportions were totaled for all dissertations (total number of readers= 50,202) across all 18 fields using the pre-defined Mendeley status categories: Ph.D. Student, Doctoral Student, Postgraduate Student, Master Student, Bachelor, Professor, Associate Professor, Senior Lecturer, Lecturer, Researcher, Librarian, Other, and Unspecified (see Appendix B). Figure 14 gives a broader view through merging readership statuses into related classes. Over two thirds (69%) of all Mendeley readers of dissertations were students (Ph.D. or doctoral: 38%, Master or Postgraduate: 23%, and Bachelor: 8%), whereas 16% were academic staff or researchers (Researcher: 10%, Professor: 2%, Associate Professor: 2%, Lecturer: 2%, and Senior Lecturer: 1%) and 15% were something else (Figure 14). Thus, students were the main Mendeley readers of dissertations. However, there are some disciplinary differences. For instance, there was a higher percentage of student readers in Economics and Business (76%) than in Health Sciences (66%) and science fields such as Earth and Environmental Sciences, Agricultural Sciences, Mathematics and Physics and Astronomy, where a greater proportion of academics read doctoral dissertations (21%-22%) than in most other fields.

In comparison to a previous result about the status of Mendeley readers of WoS articles (see Mohammadi, Thelwall, Haustein, & Larivière, 2015), a greater proportion of students read doctoral dissertations than articles in Chemistry (57% vs. 72%), Medicine (49% for articles vs. 68% for dissertations), and Physics (57% vs. 67%) but there was little difference in Social Science (65% vs. 67%) and Engineering and Technology (69% and 68.5%). The proportion of student readers of dissertations from the three fields Medicine, Health sciences and Psychology (68%) was also higher than the proportion of student readers of journal articles (56%) in another large-scale study (Haustein & Larivière, 2014) based on the four comparable disciplines (Biomedical Research, Clinical Medicine, Health and Psychology), suggesting that doctoral dissertations tend to be read more by students, perhaps to help them with their own theses. For instance, a survey in four American academic libraries showed that dissertations were more frequently borrowed by those writing theses (57%) than those writing research papers (30%) (Repp & Glaviano, 1987).

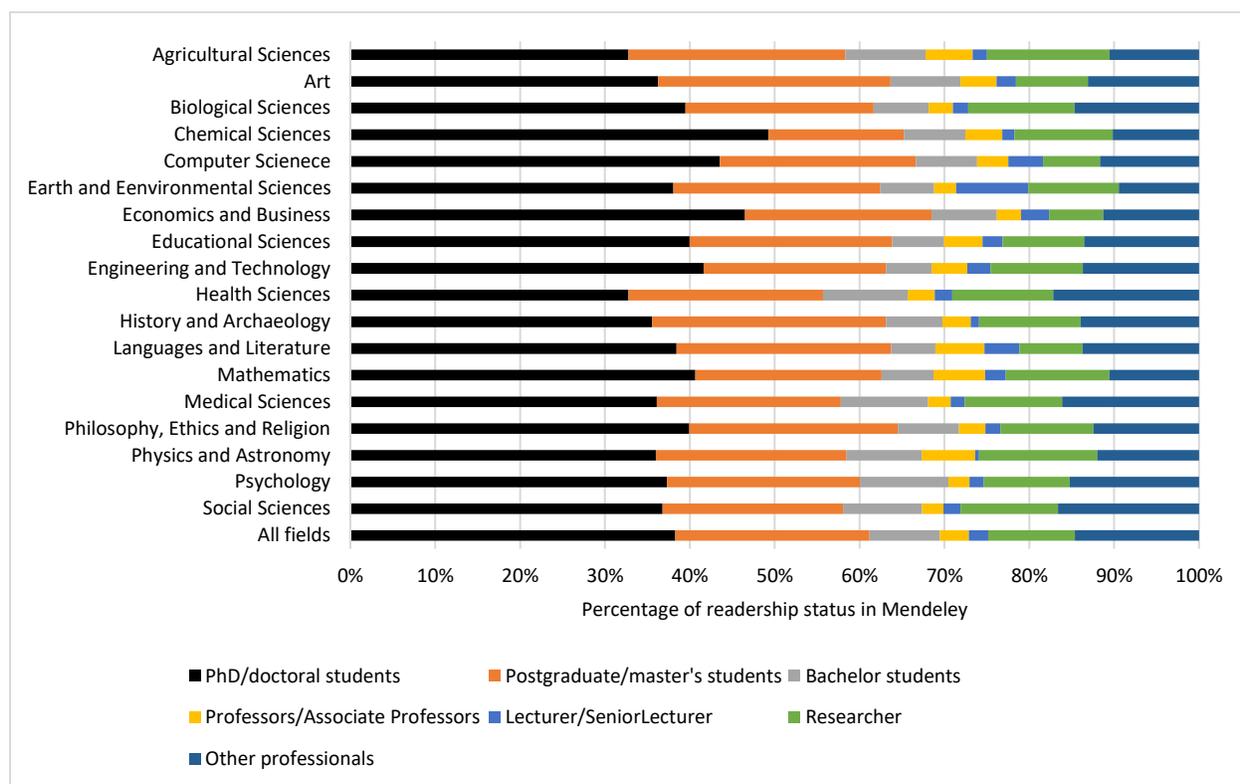

Figure 14. Reader occupational status in Mendeley for ProQuest doctoral dissertations across 18 fields during 2013-2017 (total number of readers= 50,202).

# 6  Discussion

As discussed above, the results are from 30% of all American doctoral dissertations indexed in ProQuest Dissertations & Theses from 2013-2017 investigated in this study (see discussion below). If this subset is biased in some way such as ProQuest indexing a biased subset of all US dissertations or providing a subset of dissertations to be indexed by Google Scholar (Arbor, 2017) then this will affect the graphs above. Moreover, the results for other countries may well be different and there are likely to be finer-grained disciplinary differences than analyzed here. For instance, only 6% of the ProQuest indexed doctoral dissertations 2013-2018 were in non-English languages (e.g., 56 in Italian and 7 in Russian).

Google Scholar indexes many ProQuest dissertations from other sources, such as university repositories, even when it does not return them with site:proquest.com searches. For instance, although the doctoral dissertation, "*Labor Market Shifts and their Impact on Teen Birth Rates*" published in 2017 was not searchable by Google Scholar from the proquest.com domain, Google Scholar had indexed it from the University of Chicago repository. To investigate this more, a random sample of 200 ProQuest-indexed American doctoral dissertations from 2017 were manually searched for by title in Google Scholar. Most of these provided a result in Google Scholar: 49% (98) from the .edu domain, 33.5% (67) from the proquest.com, 6% (12) from other domains (e.g., the University of California repository: https://escholarship.org/) and 11.5% (23) of the dissertations were not found in Google Scholar. This suggests that the main reason for missing many (70%) ProQuest dissertations from the original database through the semi-automatic dissertation citation search method could be that most are only searchable through their university domains in Google Scholar. For instance, 42% of the 98 doctoral dissertations found by Google Scholar from the .edu domain had other versions from proquest.com under the Google Scholar "all versions" link but these additional versions were not searchable via the site:proquest.com command. Another practical limitation is that many doctoral dissertations are only accessible in print format or may be stored in non-searchable databases or websites (e.g., http://www.dart-europe.eu). Hence, it is necessary to modify the Google Scholar citation search method to add results from other platforms and repositories. The method might be more difficult to use for repositories containing different types of academic publications (e.g., articles, books, technical reports and theses), where the *site:* commend may also retrieve many false matches from non-dissertation publications than for specific databases of theses or dissertations.

**Secondary analysis based on American university repositories:**  Because only 30% of American doctoral dissertations indexed in ProQuest were retrieved (77,884 of 264,149) for 2013-2017 by Google Scholar, an alternative method was used to estimate Google Scholar's coverage of doctoral theses from American university repositories outside the ProQuest domain. The query *"for the degree of Doctor of" site:edu author:[A-Z]* was used in Google Scholar and results were limited to 2017, finding 15,477 records from American universities. Of these, more than two-thirds (68% or 10,559 out of 15,477) matched doctoral dissertations in *the ProQuest Dissertations & Theses* database in 2017. However, less than 1% (110 out of 14,900) records matched with the initial collected dataset in 2017 from the ProQuest domain (site:proquest.com).This indicates that Google Scholar indexes a substantial number of unique ProQuest doctoral dissertations from university repositories. For instance, Google Scholar searches for 2017 American doctoral theses found 30,267 unique dissertations from either ProQuest (14,900) or American university repositories (15,367) which includes about 60% (30,267 out of 50,328) of all doctoral dissertations indexed by *the ProQuest Dissertations & Theses* database in 2017.

In answer to the first research question, citations to ProQuest doctoral dissertations can be semi-automatically extracted in a large-scale manner from the Google Scholar with accuracy high enough for citation analysis, although of its recall is lower (and difficult to exactly calculate) because not all doctoral dissertations indexed from ProQuest are associated with the ProQuest domain for Google Scholar queries. To generate a complete set of ProQuest dissertations indexed by Google Scholar it would therefore be necessary to query all other web domains for dissertations and identify those that were dual indexed with ProQuest.com. More generally, the method introduced here cannot guarantee 100% coverage of any domain because some of the dissertations in that domain may be dual indexed with another domain and only retrievable from that domain. The proportion of dissertations affected in this way seems likely to vary substantially by domain.  The method described here could be used to estimate the citation impact of dissertations in other specific databases of electronic theses, such as such the *British Library EThOS* (ethos.bl.uk), *theses.fr* (www.theses.fr) or *Theses and Dissertations Online* (www.tdx.cat) outside the USA. For instance, the query *site:ethos.bl.uk "Awarding Body: University of Cambridge"* with the author splitting technique described above can be used to obtain Google Scholar citation counts for doctoral dissertations from the University of Cambridge. The British Library theses database has indexed more doctoral dissertations from the UK (97,615) than the ProQuest database (86,664) during 2013-2017 and hence the methods can potentially be used for the impact assessment of a large number of dissertations from British universities, although it cannot guarantee to identify all dissertations because of the dual indexing problem experienced for ProQuest.

In answer to the second research question, Google Scholar citations are most numerous and therefore most useful for older dissertations (3-5 years after publication) in the social sciences, arts and humanities, but are still more numerous than Mendeley readers in science, technology and biomedical fields. Mendeley readers were more numerous and therefore potentially more useful than Google Scholar citations for recently published dissertations (1-2 years after publication), where they can be an early impact indicator.

In answer to the third research question, there were low or insignificant correlations between Google Scholar citations and Mendeley readers for doctoral dissertations. Coupled with the evidence of a substantial minority of citations from dissertations in Google Scholar, this suggests that two indicators may reflect different types of impacts, with Mendeley reflecting impact on postgraduate students to a much greater extent.

In answer to the fourth research question, Google Scholar had a high level of precision when identifying citations to doctoral dissertations (96.6%) and over half (56%) of the citations to dissertations found by Google Scholar were from journal articles, with less than a third (29%) being from other dissertations. This is four times higher than the share of citations to journal articles from dissertations (4% according to: Kousha & Thelwall, 2008), suggesting that dissertations are unusually valuable for doctoral students compared to journal articles. This may be because they are more aware of them whilst they are writing their own theses, because they read and cite their friends' dissertations, or because the extended descriptions in theses are more digestible for junior researchers. Alternatively, more senior researchers writing journal articles may tend to avoid citing theses because they take time to read. However, the practical limitation is that small number of dissertations with at least one Google Scholar citations were randomly used (n=360 or 2.3% of total results) to assess types of scholarly document cite dissertations. Hence, future studies can also investigate this using larger sample in different fields (see Appendix A).

In answer to the final research question, most readers of dissertations in Mendeley are doctoral or other postgraduate students, although the same is true for journal articles (Mohammadi, Thelwall, Haustein, & Larivière, 2015). Since Mendeley users are probably more junior than average publishing academics, this overestimates the extent to which postgraduates read dissertations. Because of this complicating factor, it is not clear whether students are more likely to read dissertations.

# 7  Conclusion

The method introduced by this paper is the first attempt to investigate the citation impact of dissertations on a large-scale through Google Scholar, making it more practical for universities, departments, employers or funders of doctoral research to assess the scholarly influence of doctoral research or junior researchers. This will not be useful for most individual researchers since most dissertations are uncited, but can help large scale evaluations of doctoral programmes or doctoral funding initiatives. Such evaluations should also take into account the impact of journal articles and other publications derived from dissertations as well as the extent of self-citations to dissertations (e.g., 22% in this study). Whilst this study has focused on ProQuest dissertations, similar methods should work for other online repositories that allow full text indexing in Google Scholar. Although Google Scholar can now be used for large scale evaluations of the impact of dissertations, a limitation is that the coverage of the method may be low. Thus, large scale evaluations would need to be carried out on whatever sample the method returns, rather than on all relevant dissertations.

Evaluators wishing to monitor the scientific performance of doctoral programs based on Google Scholar citations need to check its coverage of dissertations before performing searches. This study showed that both the proquest.com and edu domains should be considered in Google Scholar queries to identify a large number of the American doctoral dissertations. However, Google Scholar's coverage of dissertations may vary between countries and universities. For instance, the Google Scholar query site:ethos.bl.uk "University of Cambridge" retrieves about over 30,000 doctoral dissertations from University of Cambridge indexed by the British Library EThOS database. Because the University of Cambridge primarily uses the British Library EThOS database for disseminating of its Ph.D. theses (https://libguides.cam.ac.uk/theses), Google Scholar seems to be a good source to monitor the scientific impact of doctoral theses from this university. Nevertheless, some universities may have repositories of theses and dissertations with non-academic domains, and the above advice would be inadequate for these. For example, eScholarship.org is the (non-edu) scholarly publishing and repository service of the University of California. Other universities may not fully disseminate dissertations in repositories.

An alternative strategy to identify citations to dissertations from Google Scholar (suggested by a reviewer) would be to query them individually by title, if a complete list of dissertation titles for an institution is available. This would circumvent the problem of incomplete results from the *site:* command at the expense of the extra manual labour needed to submit the queries.

It is clear from the results that a minority of dissertations – at least 20% - continue to have a direct impact within academia despite the trend to publish articles during Ph.D. studies. The figures for Mendeley (16%) suggest that a higher proportion is read (excluding the candidate, supervisors and examiners), given that no more than 1 in 12 academics may be Mendeley users (Van Noorden, 2014). This must be enough to justify ongoing initiatives to digitize dissertations, especially given that this is not an onerous or resource-hungry task. Moreover, dissertations seem to be particularly useful to postgraduate students, which may be worth investigating further.

Appendix A: The number of American ProQuest doctoral dissertations identified by Google Scholar from 2013-2017 in 18 fields.

| Field | 2013 | 2014 | 2015 | 2016 | 2017 | Total |
|---|---|---|---|---|---|---|
| Agricultural Sciences | 209 | 212 | 203 | 217 | 183 | 1,024 |
| Art | 437 | 503 | 500 | 409 | 398 | 2,247 |
| Biological sciences | 1,552 | 1,580 | 1,587 | 1,471 | 1,235 | 7,425 |
| Chemical sciences | 497 | 386 | 459 | 426 | 352 | 2,120 |
| Computer science | 466 | 555 | 496 | 520 | 576 | 2,613 |
| Earth and environmental sciences | 275 | 328 | 321 | 320 | 276 | 1,520 |
| Economics, business and management | 1,030 | 1,126 | 1,085 | 1,203 | 1,118 | 5,562 |
| Educational sciences | 2,834 | 2,994 | 3,072 | 3,074 | 3,200 | 15,174 |
| Engineering and Technology | 1,465 | 1,490 | 1,571 | 1,326 | 1,415 | 7,267 |
| Health sciences | 687 | 807 | 838 | 791 | 809 | 3,932 |

| | | | | | |
|---|---|---|---|---|---|
| History and archaeology | 394 | 406 | 360 | 295 | 232 | 1,687 |
| Languages and literature | 525 | 542 | 511 | 422 | 386 | 2,386 |
| Mathematics | 388 | 371 | 414 | 394 | 431 | 1,998 |
| Medical Sciences | 992 | 1,033 | 928 | 976 | 903 | 4,832 |
| Philosophy, ethics and religion | 511 | 529 | 591 | 544 | 512 | 2,687 |
| Physics and astronomy | 514 | 503 | 603 | 538 | 579 | 2,737 |
| Psychology | 945 | 1,028 | 1,017 | 1,119 | 996 | 5,105 |
| Social sciences | 1,508 | 1,572 | 1,655 | 1,475 | 1,358 | 7,568 |
| Total | 15,229 | 15,965 | 16,211 | 15,520 | 14,959 | **77,884** |

Appendix B: Percentage of readership status in Mendeley for ProQuest doctoral dissertations across 18 fields during 2013-2017 (total number of readers= 50,202).

| Subjects of dissertations | Students | | | | | Academics and researchers | | | | | Other | | | %Total |
|---|---|---|---|---|---|---|---|---|---|---|---|---|---|---|
| | Ph.D. | Doctoral | Postgraduate | Master | Bachelor | Professor | Associate Professor | Senior Lecturer | Lecturer | Researcher | Librarian | Other | Unspecified | |
| Agricultural Sciences | 17.8% | 15% | 1.7% | 23.9% | 9.4% | 3.3% | 2.2% | 0% | 1.7% | 14.4% | 3.9% | 2.8% | 3.9% | 100% |
| Art | 27.2% | 9.1% | 2.1% | 25.3% | 8.3% | 2.5% | 1.7% | 0.3% | 1.9% | 8.5% | 3.2% | 3.2% | 6.7% | 100% |
| Biological Sciences | 24.9% | 14.6% | 4.1% | 18.1% | 6.5% | 1.5% | 1.4% | 0% | 1.8% | 12.5% | 3.5% | 3.4% | 7.7% | 100% |
| Chemical Sciences | 31.9% | 17.4% | 4.3% | 11.6% | 7.2% | 4.3% | 0% | 0% | 1.4% | 11.6% | 7.2% | 1.4% | 1.4% | 100% |
| Computer Science | 31.1% | 12.5% | 3% | 20% | 7.2% | 2.1% | 1.6% | 0.8% | 3.3% | 6.7% | 2.1% | 2.6% | 6.9% | 100% |
| Earth and Environmental Sciences | 28.1% | 9.9% | 3.8% | 20.6% | 6.4% | 0.9% | 1.7% | 0.7% | 7.8% | 10.6% | 1.9% | 3.3% | 4.3% | 100% |
| Economics and Business | 29.3% | 17.2% | 2.8% | 19.2% | 7.7% | 1.4% | 1.4% | 0.9% | 2.4% | 6.4% | 1.4% | 3.5% | 6.5% | 100% |
| Educational Sciences | 26% | 13.9% | 2% | 21.9% | 6.1% | 2.5% | 2% | 0.8% | 1.6% | 9.6% | 4% | 3.3% | 6.3% | 100% |
| Engineering and Technology | 29.5% | 12.1% | 1.5% | 20% | 5.4% | 1.7% | 2.5% | 0.3% | 2.4% | 10.9% | 4.8% | 2.5% | 6.3% | 100% |
| Health Sciences | 21.6% | 11.2% | 2.9% | 20% | 10% | 1.5% | 1.7% | 0.6% | 1.5% | 11.9% | 4.4% | 4% | 8.8% | 100% |
| History and Archaeology | 27.6% | 8% | 1% | 26.6% | 6.6% | 1% | 2.3% | 0% | 1% | 12% | 2.3% | 3.3% | 8.3% | 100% |
| Languages and Literature | 29.4% | 9% | 1.2% | 24.1% | 5.2% | 2.9% | 2.9% | 0.4% | 3.7% | 7.4% | 2.6% | 2.8% | 8.3% | 100% |
| Mathematics | 32.5% | 8.1% | 4.5% | 17.5% | 6.1% | 3.3% | 2.8% | 0.4% | 2% | 12.2% | 1.2% | 4.5% | 4.9% | 100% |
| Medical Sciences | 22.9% | 13.2% | 3.5% | 18.2% | 10.3% | 0.9% | 1.8% | 0.4% | 1.2% | 11.5% | 3.8% | 3.4% | 8.9% | 100% |
| Philosophy, Ethics and Religion | 29.6% | 10.3% | 1.7% | 22.8% | 7.2% | 0.7% | 2.4% | 0.4% | 1.4% | 11% | 0.6% | 2.1% | 9.7% | 100% |
| Physics and Astronomy | 24.4% | 11.5% | 3.3% | 19.2% | 8.8% | 3.1% | 3.1% | 0.3% | 0.1% | 13.9% | 1.9% | 2.2% | 7.8% | 100% |
| Psychology | 25.1% | 12.2% | 2.7% | 20.1% | 10.4% | 0.9% | 1.6% | 0.4% | 1.2% | 10.1% | 2.5% | 2.9% | 9.8% | 100% |
| Social Sciences | 24.5% | 12.3% | 2.8% | 18.4% | 9.3% | 0.9% | 1.6% | 0.3% | 1.7% | 11.5% | 3.2% | 3.8% | 9.6% | 100% |
| Total | 25.4% | 12.8% | 2.6% | 20.3% | 8.3% | 1.6% | 1.8% | 0.6% | 1.7% | 10.2% | 3.3% | 3.3% | 8% | 100% |